\documentclass[a4paper,11pt]{article}
\usepackage{jinstpub} % for details on the use of the package, please see the JINST-author-manual
\newcommand{\lt}{<}
\usepackage{booktabs}
\usepackage{lineno}
\title{\boldmath Electron identification and hadron discrimination using Cherenkov radiation in air and SiPMs}

\author[a,b]{A. Alici,}
\author[b,1]{F. Carnesecchi,\note{Corresponding author.}}
\author[a,b]{B.R. Achari,}
\author[a,b]{N. Agrawal,}
\author[b]{P. Antonioli,}
\author[a,b]{S. Arcelli,}
\author[a,b]{F. Bellini,}
\author[c,d]{S. Bufalino,}
\author[b]{D. Cavazza,}
\author[a,b]{L. Cifarelli,}
\author[b]{F. Cindolo,}
\author[e,b]{G. Clai,}
\author[a,b]{M. Colocci,}
\author[a,b]{F. Ercolessi,}
\author[a,b]{G. Fabbri,}
\author[b]{D. Falchieri,}
\author[d]{C. Ferrero,}
\author[f]{A. Ficorella,}
\author[c,d]{U. Follo,}
\author[g,b]{M. Garbini,}
\author[a,b]{S. Geminiani,}
\author[h]{G. Gioachin,}
\author[f]{A. Gola,}
\author[b]{D. Hatzifotiadou,}
\author[b]{A. Khuntia,}
\author[b]{A. Margotti,}
\author[a,b]{G. Malfattore,}
\author[b]{R. Nania,}
\author[b]{F. Noferini,}
\author[f]{L. Parellada-Monreal,}
\author[f]{M. Penna,}
\author[b]{O. Pinazza,}
\author[b]{R. Preghenella,}
\author[a,b]{M. Razza,}
\author[b]{R. Ricci,}
\author[b]{L. Rignanese,}
\author[c,d]{A. Rivetti,}
\author[a,b]{G. Romanenko,}
\author[a,b]{N. Rubini,}
\author[a,b]{E. Rovati,}
\author[a,b]{B. Sabiu,}
\author[b]{E. Scapparone,}
\author[a,b]{G. Scioli,}
\author[a,b]{S. Strazzi,}
\author[a,b]{S. Tomassini,}
\author[a,b]{and A. Zichichi}

\affiliation[a]{Dipartimento di Fisica e Astronomia "A. Righi", University of Bologna, viale Carlo Berti Pichat 6/2, Bologna, 40127, Italy}
\affiliation[b]{INFN, Sezione di Bologna, viale Carlo Berti Pichat 6/2, Bologna, 40127, Italy}
\affiliation[c]{Dipartimento di elettronica e telecomunicazioni, Politecnico di Torino, Corso Duca degli Abruzzi, 24, Torino, 10129, Italy}
\affiliation[d]{INFN, Sezione di Torino, Via Pietro Giuria 1, Torino, 10125, Italy}
\affiliation[e]{ENEA, Sede di Bologna, Via dei Mille 21, Bologna, 40121, Italy}
\affiliation[f]{Fondazione Bruno Kessler, Via Sommarive 18, Povo, 38123, Italy}
\affiliation[g]{Museo Storico della Fisica e Centro Studi Enrico Fermi, Via Panisperna 89 A, Roma, 10129, Italy}
\affiliation[h]{Dipartimento di Scienza Applicata e Tecnologia (DISAT), Politecnico di Torino, Corso Duca degli Abruzzi 24, Torino, 10129, Italy}

\emailAdd{francesca.carnesecchi@bo.infn.it}

\abstract{This paper presents a method to identify electrons using the Cherenkov light emitted when a charged particle travels in air and photons are detected with a Silicon PhotoMultiplier (SiPM). The analysis is based on a photon-counting approach using SPAD cells and uses data collected during a test beam at CERN PS. The results are well described by a simple Monte Carlo simulation, which further demonstrates that a very good electron identification and a strong pion/hadron rejection could be obtained over a wide momentum range.}


\arxivnumber{2601.03405}% Only if you have one

\begin{document}
\maketitle
\flushbottom

\section{Introduction}\label{sec1}

In  \cite{SiPM1, SiPM2, SiPM3, SiPM4}  it was quantitatively shown that the Cherenkov radiation produced at the passage of a charged particle through the  standard, factory-supplied protection layer of the SiPM itself can produce very high number of firing cells (SPADs with area tens of microns square). Considering a refractive index  of 1.56 and a thickness of few hundreds micron for the Silicone material of the protection layer, the Cherenkov angle reaches $\approx$50 degrees, covering a quite large area of few mm$^2$.  In these conditions  it is possible to directly detect charged particles  and reach very good efficiency and time resolution around 20 ps \cite{SiPM3,SiPM4}.  %In \cite{SiPM3} it was shown that, for protons with lower momenta, i.e.around 1 GeV/$c$,  the number of firing SPAD diminished as expected for protons approaching the Cherenkov threshold in the SiPM protection resin. 
These results opened the possibility to use just a single SiPM as an optimal detector for Time Of Flight (TOF) measurements.

The results reported in the above papers also indicate a very reduced number of firing SPADs when no protection layer is present. For such sensors, events with more than one SPAD are mainly due to cross-talk processes \cite{SiPM3,SiPM4}. 

%However, considering that particles traverse air before impinging the SiPM, the difference in the Cherenkov threshold among electrons and other particles may produce for low energy electrons an excess of photons at very small angle with a corresponding increase in the number of firing SPAD around the impact point of the particle on the SiPM. Particles like muons, pions, kaons or protons would instead not produce such effect since the Cherenkov threshold is higher than 4.0, 5.5, 17 or 32  GeV/$c$  respectively. 
However, considering particles traversing air before impinging on the SiPM, the Cherenkov effect should still occur for high beta values. This implies that for electrons, even at low energies, there could be an excess of photons emitted at very small angles. Consequently, this would lead to an increase in the number of firing SPADs around the particle's impact point on the SiPM. Particles like pions, kaons or protons would instead not produce such an effect till they reach a momentum higher than 5.7, 20 or 38 GeV/c respectively. 
As a consequence, unambiguous identification of electrons would be possible up to $\sim$5-6 GeV/$c$ by simply identifying events with SiPMs without any protection layer and a number of SPADs in excess of the cross-talk. This cell-counting procedure is based on the amplitude of the SiPM signal and its characteristic structure, in which each peak corresponds to a fired SPAD. 

%%%%%%%%%
% \begin{figure}[h!]
%        \centering%
%        \includegraphics [width=0.8\textwidth]{UPDATED_PLOT/Cherenkovangle.png}
%        \caption{Threshold Cherenkov angle %for different particles in air. } 
%          \label{fig:thCherenkov}
%\end{figure}
%%%%%%%%%
%%%%%%%%%
%  \begin{figure}[h!]
%         \centering%
%         \includegraphics [width=0.8\textwidth]{UPDATED_PLOT/T10beamcompv3.png}
%         \caption{T10 test beam at CERN:  particle composition as a function of the beam momentum \cite{vandijk}.} 
%           \label{fig:T10bcompo}
% \end{figure}
%%%%%%%%%

To test the above possibility, in this paper data collected at T10 test beam at CERN with momenta from 1.5 to 10 GeV/$c$ are used. The apparatus is the same described in \cite{SiPM3}. At lower momenta the fraction of electrons/positrons in the beam increases thus allowing to compare the SiPM (without protection) signal with the one observed for pions and protons. In Section \ref{sec2} the experimental set-up and the analysis methods are described. The results are reported in Section \ref{result} and will be compared with a simple simulation in Section \ref{simulation}. In Section \ref{optimization} a possible optimization of the parameters involved in such study (length of air traversed, different traversed gas, different SiPM technology) is discussed.

%This work is part of the R\&D studies for the TOF timing layer of the ALICE 3 \cite{ALICE3} experiment at the LHC.

\section{Experimental setup and analysis method}\label{sec2}

\subsection{Detectors} 
\label{sec:detectors}
For the present study NUV-HD-LFv2 SiPMs produced by Fondazione Bruno Kessler (FBK) were used  \cite{2020Mazzi, Altamura, ALTAMURA2023}.
These detectors have 
%are the same used in \cite{SiPM3} with 
an active area of 3.20$\times$3.12 mm$^2$ filled with 6200 SPADs with square pixel pitch of 40 $\mu \text{m}$, 83$\%$ fill factor and breakdown voltage V$_{bd}$ = 32.2 $\pm$0.1 V \cite{2019gola,Gundacker_2023}.
%\cite{SiPM1}.
%The detectors were produced with standard protection layers of 1.0 and 1.5 mm silicone resin (refraction index 1.5)  or without any protection layer.
In this paper only sensors without any protection layer are studied.
%We remind that, since the sensor itself is 550 $\mu$m thick, the two thicknesses correspond to an effective protection layer on top of the sensor of 450  and 950 $\mu$m  respectively for the 1.0 and 1.5 mm. 

\subsection{Beam test setup} 
\label{subsec:tb}
The SiPMs were tested at the CERN PS T10 particle beam in the momentum range 1.5-10~GeV/$c$. The beam composition, evaluated via simulation and data,  is reported in fractions with few \% uncertainties, in \cite{vandijk}: while at high momenta the beam is mainly made of  protons and $\pi^+$, at lower energies the fraction of positrons increases and becomes dominant. In the following, we will neglect the charge, which is uninfluential for our purposes, and use generically the terms pions and electrons.

The apparatus was made of a telescope with four sensors: two SiPMs under test and two LGAD detectors (1x1 mm$^2$ area and 35 \textmu  m or 25 \textmu  m thick sensors) \cite{LGAD}. The latter were used as timing reference and were placed at the beginning and at the end of the telescope to define, through their coincidence, a trigger for the beam particles. The four sensors were placed at a distance of about 7 cm one from the other. Each sensor was mounted on a remotely controlled movable frame capable of positioning with 10 $\mu$m precision in both directions perpendicular to the beam axis, ensuring accurate alignment with the beam line.
%For the SiPM without protection layer the copper Faraday cage, usually used to shield the front-end electronics, had a hole (with area around 35 mm$^2$) to allow the photons produced in the preceding 7 cm of air to reach the sensor. 
One of the two DUT SiPMs was without the protection layer and was used for this analysis. For this SiPM the copper Faraday cage, normally used to shield the front-end electronics at about 6 mm distance from the SiPM surface, had a hole (with area around 35 mm$^2$) to allow the photons produced in the preceding 7 cm of air to reach the sensor. The SiPM signals were coupled to a customized front-end with X-LEE amplifiers\footnote{\href{https://www.minicircuits.com/pdfs/LEE-39+.pdf}{https://www.minicircuits.com/pdfs/LEE-39+.pdf}}  with a total gain factor of about 40 dB. The set-up was inserted into a light-tight box and is the same as described in details in \cite{SiPM3}. All sensors were operated at ambient temperature. Waveforms from all four sensors were recorded via a Lecroy Wave-Runner 9404M-MS digital oscilloscope.

\subsection{Analysis method}
\label{subsec:analysis}

\begin{figure}[h!]
        \centering
        \includegraphics [width=1.00\textwidth]{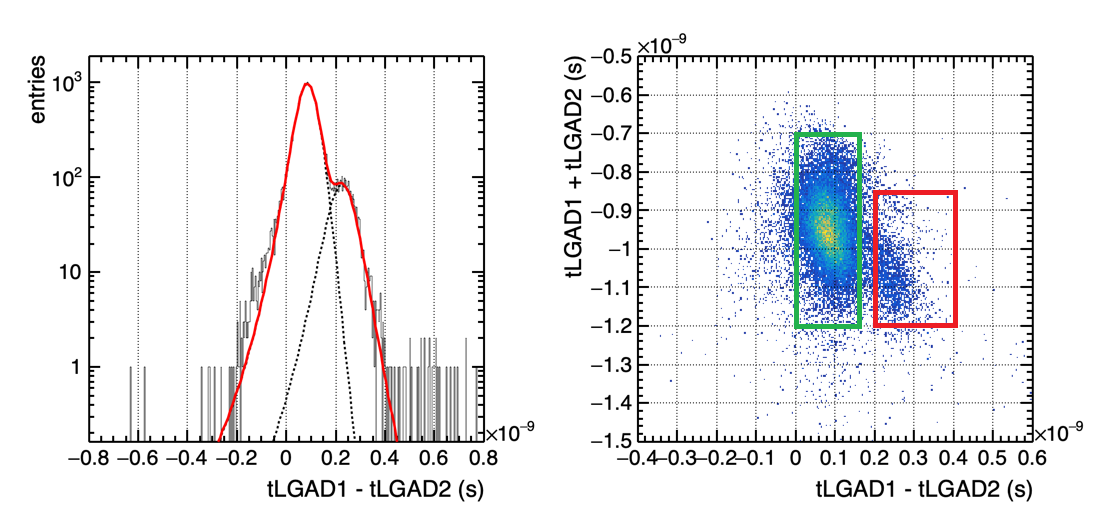}
        \caption{Left panel: time of flight of the beam particles computed as the time difference between the two LGADs in the telescope at 1.5 GeV/$c$. The fits are performed with two q-Gaussian, the right one relative to the proton content and the left one relative to pions/electrons. Right panel: LGAD time sum versus time difference distribution with the box cut for protons (red) and electron/pions (green).} 
          \label{fig:LGAD_tof}
\end{figure}

The analyzed data were collected with a beam momentum of 1.5 GeV/$c$ (a second sample at 10 GeV/c was also collected to check the simulation). The reason for this choice is twofold. First, it enhances the electron component in the beam, which is composed of approximately 60\% electrons, 28\% pions, and 12\% protons \cite{vandijk}. Notice that these fractions are estimated with few \% uncertainties and such variations do not affect the results reported in Section 3. Second, it enables the separation of electrons and pions from protons by exploiting the time of flight between the signals from the two LGAD detectors placed at the ends of the telescope, as visible in Fig.~\ref{fig:LGAD_tof} left (more details in \cite{SiPM3}). In the figure a q-Gaussian fit \cite{q-gauss} was applied on both distributions.

At this beam momentum, electrons are the only particles that produce Cherenkov radiation in the air gap between sensors in the telescope. Electrons and pions cannot be distinguished using the LGADs' time information; however, their charge spectra can still be determined. The proton charge spectrum is used as a proxy for pions normalized to the expected pion fraction \cite{vandijk}, since neither pions nor protons are able to generate Cherenkov radiation in air.

The protons sample has been selected with rectangular cuts on the LGAD time sum versus time difference distribution (Fig.~\ref{fig:LGAD_tof} right) as $0.2 < t_{LGAD1} - t_{LGAD2} < 0.4$~ns and $-1.2 < t_{LGAD1} + t_{LGAD2} < -0.85$~ns, while for electrons and pions the cuts are $0 < t_{LGAD1} - t_{LGAD2} < 0.16$~ns and $-1.2 < t_{LGAD1} + t_{LGAD2} < -0.7$~ns. With these cuts the fraction of protons remaining in the electron/pion sample is below 1\%, while the electron/pion contamination in the proton sample is $\approx$  4-5\%.

The SiPM signals used in the analysis are those within a window of $\pm$ 2 ns from the (t$_0$) trigger given by the coincidence of the two LGAD sensors.
The baseline of the acquired waveforms is subject to fluctuations due to the electromagnetic noise present in the experimental area, the dark count rate of the SiPM and high particle rate inducing overlapping signals. 
%To mitigate this, a time window was defined in the region preceding the signal window  between -10 ns and -2 ns from the trigger time t$_0$ within which the mean and RMS of the signal amplitude were calculated. 
To address this, a time window was defined in the interval preceding the signal window, from –10 ns to –2 ns relative to the trigger time t$_0$, within which the mean and RMS of the signal amplitude were calculated.
To remove events affected by spurious signals, noise, or after-pulses from previous events, all events with a baseline RMS greater than 0.75 mV were rejected (14\% of the total number of events). 
%The baseline was then recalculated within the same time window using only the selected events, and subsequently subtracted from the signal event by event.

\section{Results}\label{result}

The data were collected at p = 1.5 GeV/$c$ supplying the SiPM with an overvoltage (OV) of 2 V. This setting, although not optimal to distinguish between electron and hadron signals, was chosen to prevent signal saturation on the vertical scale of the oscilloscope and to ensure a good separation of signals corresponding to different numbers of fired SPADs per event.
 \begin{figure}[h!]
        \centering
        \includegraphics [width=0.45\textwidth]{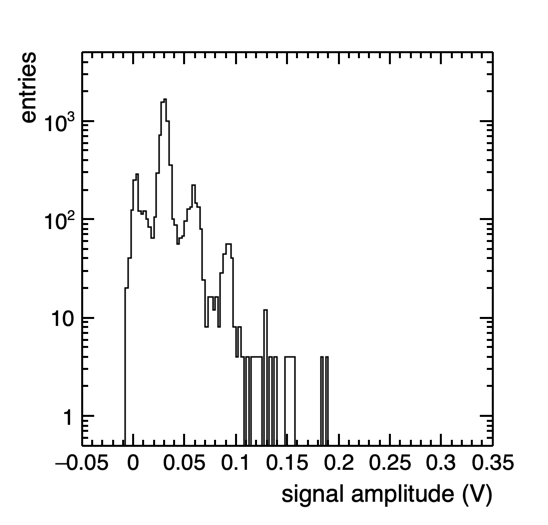}
        \includegraphics [width=0.45\textwidth]{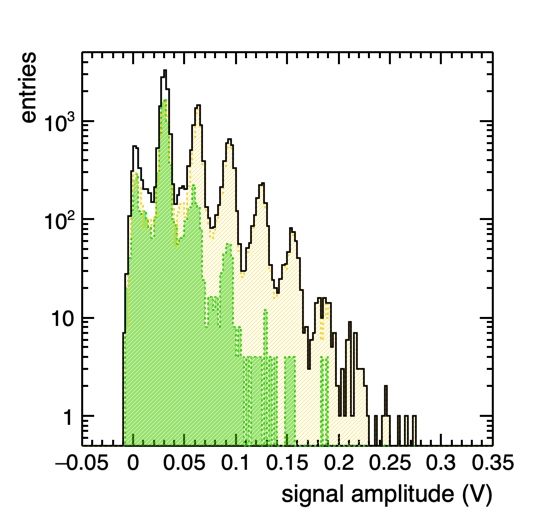}
        \includegraphics [width=0.45\textwidth]{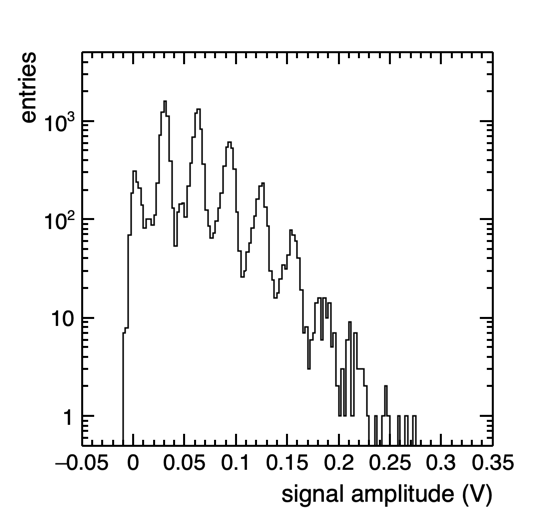}
        \caption{Real data at 1.5 GeV/c. Top left: signal amplitude for the proton sample selected. Top right: amplitude for pions and electrons with the estimated fraction of pions (green dotted area) and electrons (orange dashed area). Bottom: the estimated signal amplitude for electrons once the pion contribution (distribution expected to be identical to the protons one) is subtracted.\\
        NUV-HD-LFv2, 3.20$\times$3.12 mm$^2$, OV = 2V, thickness$_\text{air}$ = 7 cm.} 
          \label{fig:measured_spectra}
\end{figure}
The measured amplitudes for the proton sample is reported in Fig.~\ref{fig:measured_spectra} top-left: as expected, it is peaked at one SPAD ($\sim$31~mV), with the entries at 0 mV (0 SPAD) compatible with the value of the fill factor of the sensor and with peaks at 2 and 3 SPADs due to crosstalk and possible electron contamination ( see \ref{subsec:analysis} ). %The amplitude distribution for pions is expected to be identical to that of protons, while the one of electrons can be derived by taking the amplitude distribution of events identified as electrons/pions and subtracting the amplitude distribution of protons, normalized to the expected pion fraction in this sample which is estimated at 32\%\cite{vandijk}.
The amplitude distribution for pions is expected to be identical to that of protons. The electron distribution is then obtained by taking the amplitude distribution of events identified as electron/pion candidates and subtracting the proton distribution, normalized to the expected pion fraction of 32\%\cite{vandijk}.
The pion/electron spectrum is reported in Fig.~\ref{fig:measured_spectra} top-right with the estimated fraction of pions and electrons. The electron spectrum, after the pion subtraction, is instead reported in Fig.~\ref{fig:measured_spectra} bottom. 

Despite the low overvoltage, the two distributions are clearly distinguishable. Since a single SPAD signal is measured to have an amplitude of 31 mV, selecting tracks with $\geq$ two SPADs (threshold set at $\sim$ 50 mV) one gets a hadron rejection of $\approx$ 85\% with an electron selection efficiency of $\approx$ 57\%.

%\subsection{Electron, proton and pions identification}
%\label{subsec:elident}

\section{Comparison with simulation}
\label{simulation}

The results reported in the previous section rely mainly on the purity of the selected proton sample and on an accurate knowledge of the actual composition of the particle beam. To estimate the impact of these experimental uncertainties
%, which are in part beyond our full control, 
the results were cross-checked against the predictions of a toy Monte Carlo (MC) simulation to assess their robustness.
The simulation works as follows: two hits are randomly generated on the two LGADs and the track is then propagated to the SiPM. The number of Cherenkov photons created in the air volume upstream the sensor has been calculated based on the Frank-Tamm formula \cite{FrankTamm}. The detection efficiency for the charged particle was assumed equal to the fill factor according to the manufacturer's specifications, as well as the crosstalk and the PDE as a function of the photon wavelength and SiPM overvoltage. The amplitude distribution is obtained by multiplying the number of fired cells by the average amplitude of 1 SPAD signal assumed as 31 mV according to the acquired data. A gaussian spread of 3.3 mV was applied driven by data observation. Given the binary nature of the SPAD response, multiple photons hitting the same pixel are counted as one. %A still-not-fully-satisfactory modeling of the noise is applied to the simulated data based on the measured RMS of the baseline which, however, does not appear to have a significant effect on data reproducibility.
A still-not-fully-satisfactory modeling of the noise is applied to the simulated data based on the measured RMS of the baseline. However, no explicit simulation of the electronic jitter or of the detailed baseline fluctuations described in Sec.~\ref{subsec:analysis} is included, which may affect in particular the population of events between peaks and the 0-signal region.

 \begin{figure}[h!]
        \centering
        \includegraphics [width=1.0\textwidth]{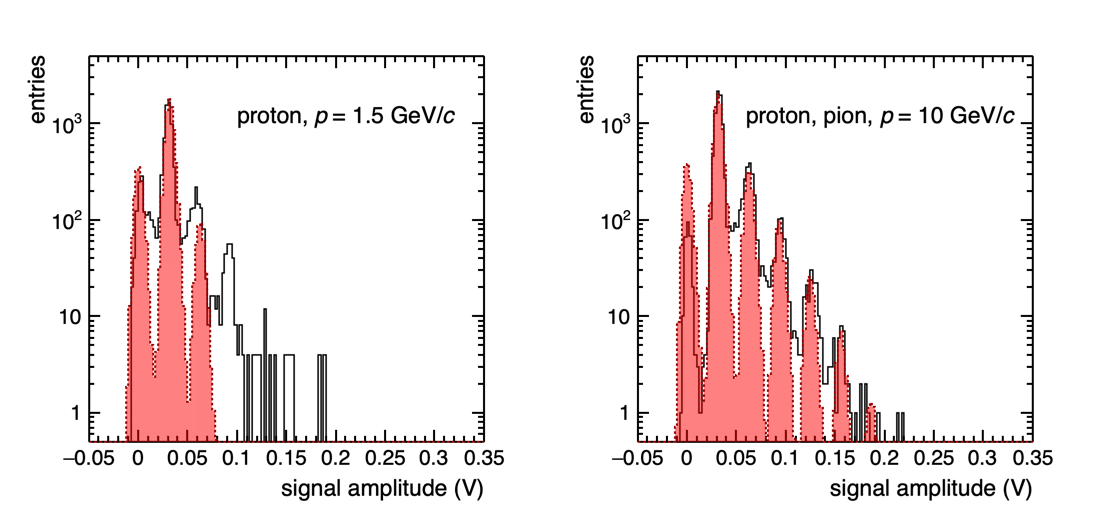}
        \caption{Amplitude distribution  of data (full line) and MC simulation (red area) for protons at 1.5 GeV/$c$ (left) and protons/pions at 10 GeV/$c$ (right).NUV-HD-LFv2, 3.20$\times$3.12 mm$^2$, OV = 2V, thickness$_\text{air}$= 7 cm.} 
          \label{fig:sim_results}
\end{figure}

 \begin{figure}[h!]
        \centering
        \includegraphics [width=0.5\textwidth]{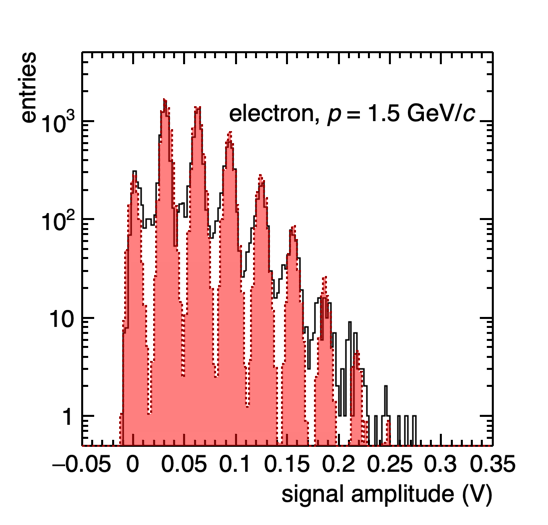}
        \caption{Amplitude distribution  of data (full line) and MC simulation (red area) for electrons at 1.5 GeV/$c$.
        NUV-HD-LFv2, 3.20$\times$3.12 mm$^2$, OV = 2V, thickness$_\text{air}$= 7 cm.} 
          \label{fig:sim_results_2}
\end{figure}

A qualitative comparison between the measured spectra and MC predictions is shown in Fig.~\ref{fig:sim_results}. In the left pad, the measured amplitude spectrum for protons with a momentum of 1.5 GeV/$c$ is compared with simulations. Here and in the following, the two histograms under comparison are normalized to the same number of entries. The measured spectrum shows a small peak at 3 SPADs which is not predicted by the toy MC. This discrepancy ($\sim 3\%$ of events with more than 2 SPADs) mainly arise from a small electron contamination in the proton sample selected by LGAD timing information, $\sim 2-3\%$; the remaining part may originate from events in which crosstalk affects more than one pixel, which is not included in the MC, or from particle near the border of two SPADs. 
In the same figure the right panel shows a comparison between measured amplitude spectra and simulations at a beam momentum of 10 GeV/$c$. At this momentum, the beam consists mainly of pions (30\%) and protons (70\%), with pions producing Cherenkov radiation in air, while protons remain below threshold\footnote{As from \cite{vandijk}, the fractions used in this comparison are a simplification and the uncertainties of few \% may affect especially the null signal }. %In this case, the agreement between the two distributions is qualitatively good and the problems noticed in the previous plot are less evident due to the higher number of SPADs firing. 
In this case, the agreement between the two distributions is overall satisfactory in terms of peak positions and relative populations, although discrepancies in the inter-peak and 0-signal regions are still visible.
Finally, Fig.~\ref{fig:sim_results_2} compares the measured and simulated amplitude spectra for electrons at 1.5 GeV/$c$ : here too, the comparison shows good agreement. 

%It should be noticed that the Toy MC does not include any electronic jitter or the baseline simulation described in \ref{subsec:analysis} which affect in particular the shape in between peaks and the null signal. This comparison with the Toy MC is qualitative and is used only to support the interpretation of the experimental findings. A more detailed simulation should be performed to obtain quantitative evaluations. 
It should be stressed that this toy MC comparison is intended to provide a qualitative support to the interpretation of the experimental observations. A more detailed simulation, including a refined description of the electronic noise and baseline response, would be required for a fully quantitative assessment.

\section {A simple optimization of the key parameters}\label{optimization}

Before using the toy MC to obtain an approximate estimate of the potential performance of such a system, a number of parameters were optimized. It was observed that the SiPM coverage has a significant impact, with larger covered areas able to collect more photons produced by the incident charged particle. 
%Furthermore, reducing the cell size, within the range tested, were not found to significantly change the performance. 
No significant changes in performance were observed reducing the cell size, within the range tested.
A comparison was therefore made between SiPMs with active areas of 3.20 $\times$ 3.12 mm$^2$, as the one used in the test beam data,  and a hypothetical  6 $\times$ 6 mm$^2$, both featuring square SPADs with a pitch of 40 \textmu m. To determine the optimal thickness of the air layer in front of the sensor, the simulation considers particles impinging the SiPM in the center. The over-voltage was set to  6 V to profit of the improved PDE.
%reproduced the configuration used in the test beam data collection, namely with two 1 mm$^2$ area LGAD sensors defining the trigger. In this setup, all charged particles pass through a region near the center of the SiPM, so edge effects are small or even negligible, particularly in the case of the 6 $\times$ 6 mm$^2$ sensor.

\begin{figure}[h!]
        \centering
        \includegraphics [width=0.8\textwidth]{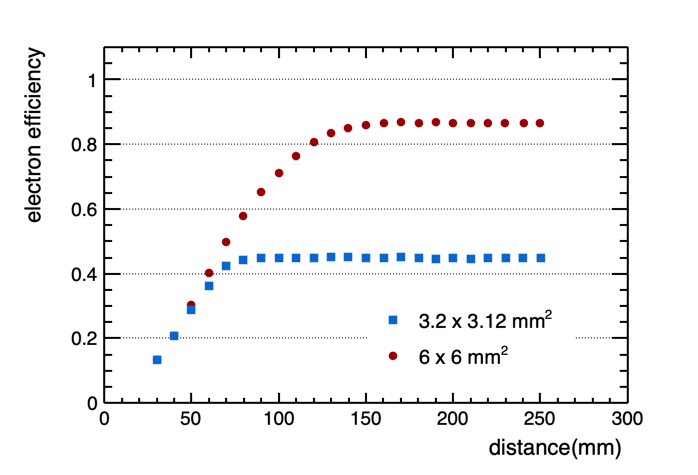}
        \caption{Electron detection efficiency versus the distance in air traversed by the particle. Electrons of 1.5 GeV/$c$ and two different SiPMs surfaces are considered.  The toy Monte Carlo parameters used are:
        NUV-HD-MT, OV = 6 V, Thr = 3 SPAD.
        }
          \label{fig:gas_thickness}
\end{figure}

The optimization was carried out using the plot shown in Fig.~\ref{fig:gas_thickness}, where the electron efficiency was calculated considering particles with 1.5 GeV/$c$ of momentum and by selecting only events in which three or more SPADs were fired. As can be seen, while below 4 cm the curves relative to the two different size SiPM overlap, at higher distances the larger-area SiPM exhibits a significantly higher efficiency
%. As mentioned, 
and this is due 
%both 
to the larger surface area, which allows more Cherenkov photons to be collected.
%, and to the fact that, given the trigger definition, edge effects are essentially negligible. 
The figure also indicates that for the 6 $\times$ 6 mm$^2$ sensor beyond 15 cm of air no appreciable variations in the number of collected photons are observed  due to the limited size of the simulated sensor. 
%Therefore, in the following simulations, this value will be used for the air layer in front of the sensor.
In the following simulations, the response of the larger SiPM will be considered using a distance of 15 cm of traversed air. Notice that the goal of this paper is to investigate the method to identify particles via the SiPM response and there is no attempt to optimize the detector configuration as it should be done for specific applications.

Since electron identification is performed by considering signals with an amplitude higher than the cross-talk value, it is clear that this value represents a fundamental parameter. Recently, the FBK NUV-HD SiPM technology has been upgraded with metal-filled deep trench isolation (NUV-HD-MT \cite{NUV-HD-MT}) between the SiPM cells, which enables an almost complete suppression of internal optical cross-talk. In the simulation, therefore, in order to select the optimal configuration for the proposed detector, the cross-talk values measured by FBK for this new technology were used. The PDE also impacts performance, as it parameterizes the probability of detecting the produced Cherenkov photons. Assuming the validity of the simulation also at overvoltages higher than 2 V, a value of 6 V was chosen to profit of the higher PDE ( from about 40\% to 60\%).
%In the results presented below, an overvoltage of 6 V was chosen; it is an intermediate value ???? that ensures a sensor response not too different from the one experimentally measured at OV = 2 V, while at the same time providing a higher PDE. 
The cross-talk and PDE values corresponding to OV = 6 V were taken from \cite{NUV-HD-MT}: possible uncertainties related to the extrapolation from 2 to 6 V overvoltage would require a careful check with experimental data in future beam test campaigns.

\begin{figure}[h!]
        \centering
        \includegraphics [width=0.75\textwidth]{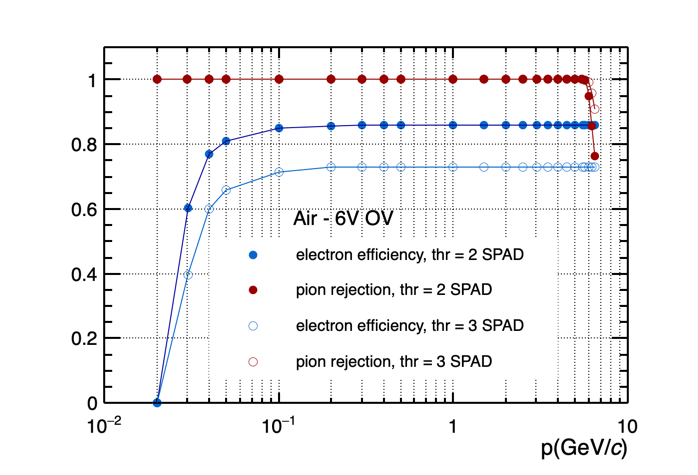}
        \includegraphics [width=0.75\textwidth]{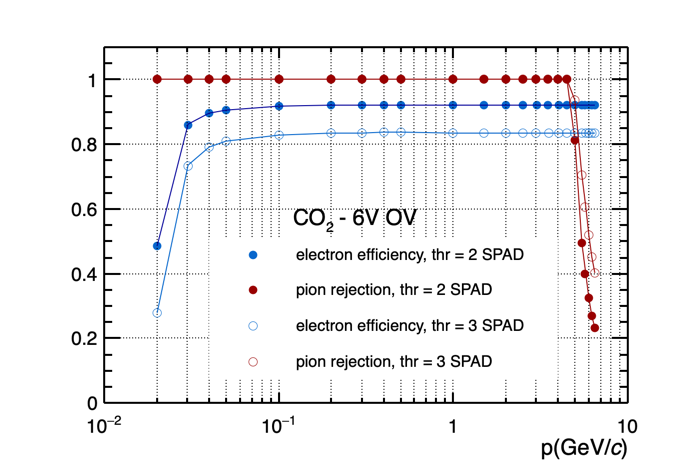}
        \caption{MC simulation of electron efficiency and pion rejection versus the momentum for two different selection cuts (number of fired SPADs $\ge$ 2 and $\ge$ 3). The toy Monte Carlo parameters used are:
        NUV-HD-MT, 6$\times$6~mm$^2$, OV = 6V, thickness$_{\text{air}/\text{CO}_2}$= 15 cm.
        } 
          \label{fig:separation}
\end{figure}

Figure \ref{fig:separation} shows electron efficiency and pion rejection as a function of the incoming particle's momentum from the toy MC simulations, for two different selection cuts (number of fired SPADs $\ge$ 2 and $\ge$ 3). In addition to the case with air (n = 1.00029) as radiator, CO$_2$ (n = 1.00045) was also considered, as it is an inert, non-flammable gas with low environmental impact and low cost; moreover, its use could be advantageous in configurations where the system requires cooling below the dew point. For both radiator gases, a pion rejection close to 100\% with an electron efficiency larger than $\sim$ 85\% looks achievable (larger than $\sim$ 90\% with CO$_2$). The choice of radiator gas leads to a shift in the results as a function of momentum. When air is used, a separation between electrons and charged hadrons can be achieved from about 50 MeV/$c$ up to $\sim$ 6 GeV/$c$, whereas CO$_2$ allows an operation down to momentum values below 30 MeV/$c$ and up to 4 GeV/$c$.

Given the threshold nature of the Cherenkov effect and the binary response of the SPADs, it is possible to define momentum regions where, although the identification of individual particles is not always feasible, an appropriate selection based on the number of fired pixels allows discrimination among different particle species.
%An example of such usage is reported in Tab.~\ref{tab:separation}, where a SiPM of 6 $\times$ 6 mm$^2$ of area, 15 cm of air as radiator and OV = 6 V was considered.

The results, obtained for a 6 × 6 mm$^2$ SiPM with 15 cm of air as radiator and an overvoltage of 6 V,  show that the number of fired SPADs can serve as a practical and tunable handle for particle discrimination. In particular, accordingly to Table~\ref{tab:separation},  low-momentum electrons and very high-momentum protons can each be effectively separated, while at the same time it is possible to set high rejection criteria for the different species in other momentum ranges.  

\begin{table}[h!]
\caption{Particle selection in different momentum ranges based on the number of fired SPADs for a 6~$\times$~6~mm$^2$ SiPM and 15 cm of air gap.
\\
} 
\label{tab:separation}
\centering
\setlength{\tabcolsep}{2pt}
\begin {tabular}{ c c c c c c }
\toprule
\textbf{p (GeV/$c$)} & \textbf{selection} & \textbf{$e$ efficiency}  & \textbf{$\pi$ rejection} & \textbf{$K$ rejection} & \textbf{$p$ rejection} \\
\midrule
0.05 -- 6  & nSPAD $\ge$ 3 & 0.86 & 0.95 & 1 & 1\\
\midrule
\textbf{p (GeV/$c$)} & \textbf{selection} & \textbf{$e$ rejection}  & \textbf{$\pi$ rejection} & \textbf{$K$ efficiency} & \textbf{$p$ efficiency} \\
\midrule
6 -- 21  & nSPAD $\lt$ 3 & 0.86 & 0.84 & 0.96 & 1\\
\midrule
\textbf{p (GeV/$c$)} & \textbf{selection} & \textbf{$e$ rejection}  & \textbf{$\pi$ rejection} & \textbf{$K$ rejection} & \textbf{$p$ efficiency} \\
\midrule
21 -- 40  & nSPAD $\lt$ 3 & 0.86 & 0.86 & 0.78 & 0.96\\
\bottomrule
\end{tabular}
\end{table}

%It is interesting to note that a configuration with two layers of SiPMs, one without and one with protection layer, placed in a structure allowing the particles to travel a path of 15 cm in air (or CO$_2$), would satisfy both requests of electron identification and TOF measurements.
Starting from the excellent timing capability of SiPM with a protection layer reported in \cite{SiPM3}, the approach discussed in the present paper suggests that configurations with two consecutive SiPMs, one without and one with protection layer, with an air gap in front of the first one, could be used to simultaneously support electron identification and TOF measurements, potentially covering PID and particle rejection in different momentum ranges. %In particular, a setup with two layers, one without and one with protection layer, allowing particles to travel a path of 15 cm in air (or CO$_2$) satisfies both requirements.
%These results naturally point to the two-layer configuration as a promising solution, the potentiality of a single-layer SiPM with protection resin and an air gap could be also investigated
%In this framework, it could also be interesting to investigate the possibility to obtain similar performances using only a SiPM with protection layer and an air gap in front: a very challenging configuration, but certainly more simple.
%while a single-layer SiPM incorporating both a protective resin and an air gap could potentially achieve the same functionality within a single sensor, an option that future investigations may explore ????.

%Such configurations, although of interest in high energy experiments, is at present mainly limited by the radiation hardness of the SiPMs in their operational environment, which remains an important aspect to address in future studies.
Such configurations would be of great interest in particle physics experiments, including space applications. However, the limited radiation hardness of the SiPMs especially in the harshest radiation environments as HL-LHC, remains an important aspect to be addressed.

\section{Conclusions}\label{conclusion}

In this paper we have studied the possibility to use SiPMs,  without coupling to scintillator and without any protection layer, to identify electrons simply using the excess of firing cells (SPADs) produced by the Cherenkov photons emitted by the electron when traversing an air region just before the sensor. Test beam data with 1.5 GeV/$c$ particles results indicate a very good electron identification and a corresponding very high pion rejection. Data are well reproduced with a simple Monte Carlo simulation which includes the photons produced by the Cherenkov effect in air and the SiPM detection parameters. 

The MC has also been used for a first simple optimization of the detector parameters indicating, for the case of 15 cm traversed air and a SiPM of 6 $\times$ 6 mm$^2$ area, the possibility of an electron identification above 85\% in the region 0.05 to 6 GeV/$c$ with a corresponding pion rejection above 96\%. This identification technique allows also to use the SPADs counting method to discriminate, in specific very high momentum regions, different particle species (K and p).

%These results demonstrate that a simple layer of SiPMs can indeed be used to identify or discriminate particles in a very large momentum range. Together with the already published results on timing performances with a protection layer, they underlie the wide range of applications of SiPM not only as sensors but also as particle detector themselves.

Together with the already published results on timing performances of SiPMs with a protection layer \cite{SiPM3,SiPM4}, the present results underline that the Cherenkov light produced in the simplest radiators in front of the SiPM, i.e. the factory-supplied silicone protective resin or the air, allow these photo-sensors to become themselves optimal detectors for charged particle identification both via time of flight or via photon counting measurements.

\section*{Declarations}
This project has received funding from INFN, FBK and the European Unions Horizon Europe research and innovation programme under grant agreement No 101057511.
The authors received research support from institutes as specified in the author list below the title. \\

\acknowledgments
%The present work is mainly due to the efforts and ingenuity of Andrea Alici who, unfortunately, passed away few days before its publication. His memory will live on with his colleagues.
The present work is mainly the result of the efforts and ingenuity of Andrea Alici, who unfortunately passed away just a few days before its publication. His memory will live on among his colleagues.
%This is the most common positions for acknowledgments. A macro is
%available to maintain the same layout and spelling of the heading.

% Bibliography

%% [A] Recommended: using JHEP.bst file
\bibliographystyle{JHEP}
\bibliography{biblio.bib}

%% or
%% [B] Manual formatting (see below)
%% (i) We suggest to always provide author, title and journal data or doi:
%% in short all the informations that clearly identify a document.
%% (ii) please avoid comments such as "For a review'', "For some examples",
%% "and references therein" or move them in the text. In general, please leave only references in the bibliography and move all
%% accessory text in footnotes.
%% (iii) Also, please have only one work for each \bibitem.

%\begin{thebibliography}{99}

%\bibitem{a}
%Author,
%\emph{Title},
%\emph{J. Abbrev.} {\bf vol} (year) pg.

%\bibitem{b}
%Author,
%\emph{Title},
%arxiv:1234.5678.

%\bibitem{c}
%Author,
%\emph{Title},
%Publisher (year).

%\end{thebibliography}
\end{document}